\input{aipcheck}

\documentclass[
    ,final            
  ]
  {aipproc}

\layoutstyle{6x9}

\begin{document}

\title{The frontier of darkness: the cases of GRB\,040223, GRB\,040422, GRB\,040624}


\author{P. D'Avanzo}{
  address={INAF, Osservatorio Astronomico di Brera, via E. Bianchi 46, I-23807 
  Merate (LC), Italy}
  ,altaddress={Universit\`a Insubria, Dipartimento di Fisica e Matematica, via 
  Valleggio 11, I-22100 Como, Italy}
}

\author{P. Filliatre}{
  address={Laboratoire Astroparticule et Cosmologie, UMR 7164, 11 place Marcelin 
  Berthelot, F-75231 Paris Cedex 05, France
}
  ,altaddress={Service d'Astrophysique, CEA/DSM/DAPNIA/SAp, CE-Saclay, Orme des 
  Merisiers, B\^{a}t. 709, F-91191 Gif-sur-Yvette Cedex, France
} 
}

\author{P. Goldoni}{
  address={Laboratoire Astroparticule et Cosmologie, UMR 7164, 11 place Marcelin 
  Berthelot, F-75231 Paris Cedex 05, France
}
  ,altaddress={Service d'Astrophysique, CEA/DSM/DAPNIA/SAp, CE-Saclay, Orme des 
  Merisiers, B\^{a}t. 709, F-91191 Gif-sur-Yvette Cedex, France
} 
}

\author{L. A. Antonelli}{
  address={INAF, Osservatorio Astronomico di Roma, via Frascati 33, Monteporzio 
  Catone, I-00040 Roma, Italy}
}

\author{S. Campana}{
  address={INAF, Osservatorio Astronomico di Brera, via E. Bianchi 46, I-23807 
  Merate (LC), Italy}
}

\author{G. Chincarini}{
address={Universit\`{a} degli studi di Milano-Bicocca, Dipartimento di Fisica, 
Piazza delle Scienze 3, I-20126 Milano, Italy}
}

\author{S. Covino}{
  address={INAF, Osservatorio Astronomico di Brera, via E. Bianchi 46, I-23807 
  Merate (LC), Italy}
}

\author{A. Cucchiara}{
  address={Department of Astronomy and Astrophysics, Pennsylvania State 
  University, 525 Davey Laboratory, University Park, PA 16802}
}

\author{M. Della Valle}{
  address={INAF, Osservatorio Astrofisico di Arcetri, largo E. Fermi 5, I-50125 
  Firenze, Italy}
}

\author{A. De Luca}{
  address={INAF - Istituto di Astrofisica Spaziale e Fisica Cosmica di Milano, 
  via E. Bassini 15, I-20133 Milano, Italy}
}

\author{S. Foley}{
  address={Department of Experimental Physics, University College Dublin, 
  Dublin 4, Ireland}
}

\author{D. Fugazza}{
  address={INAF, Osservatorio Astronomico di Brera, via E. Bianchi 46, I-23807 
  Merate (LC), Italy}
}

\author{N. Gehrels}{
  address={NASA Goddard Space Flight Center, Code 661, Greenbelt, MD 20771}
}

\author{D. G\"otz}{
  address={INAF - Istituto di Astrofisica Spaziale e Fisica Cosmica di Milano, 
  via E. Bassini 15, I-20133 Milano, Italy}
}

\author{L. Hanlon}{
  address={Department of Experimental Physics, University College Dublin, 
  Dublin 4, Ireland}
}

\author{G. L. Israel}{
  address={INAF, Osservatorio Astronomico di Roma, via Frascati 33, Monteporzio 
  Catone, I-00040 Roma, Italy}
}

\author{D. Malesani}{
  address={International school for advanced studies (SISSA/ISAS), via Beirut 
  2-4, I-34014 Trieste, Italy}
}

\author{B. McBreen}{
  address={Department of Experimental Physics, University College Dublin, 
  Dublin 4, Ireland}
}

\author{S. McBreen}{
  address={Astrophysics Missions Division, Research Scientific Support 
  Department of ESA, ESTEC, Noordwijk, The Netherlands}
}

\author{S. McGlynn}{
  address={Department of Experimental Physics, University College Dublin, 
  Dublin 4, Ireland}
}

\author{S. Mereghetti}{
  address={INAF - Istituto di Astrofisica Spaziale e Fisica Cosmica di Milano, 
  via E. Bassini 15, I-20133 Milano, Italy}
}

\author{L. Moran}{
  address={Department of Physics \& Astronomy, University of Southampton, 
  Southampton, SO17 1BJ, United Kingdom}
}

\author{J. A. Nousek}{
  address={Department of Astronomy and Astrophysics, Pennsylvania State 
  University, 525 Davey Laboratory, University Park, PA 16802}
}

\author{R. Perna}{
  address={Department of Astrophysical and Planetary Sciences, University of 
  Colorado at Boulder, 440 UCB, Boulder, CO, 80309, USA}
}

\author{L. Stella}{
  address={INAF, Osservatorio Astronomico di Roma, via Frascati 33, Monteporzio 
  Catone, 00040 Rome, Italy}
}

\author{G. Tagliaferri}{
  address={INAF, Osservatorio Astronomico di Brera, via E. Bianchi 46, I-23807 
  Merate (LC), Italy}
}

\begin{abstract}
Understanding the reasons for the faintness of the optical/near-infrared afterglows of the
so-called dark bursts is essential to assess whether they form a subclass of GRBs,
and hence for the use of GRBs in cosmology. With VLT and other
ground-based telescopes, we searched for the afterglows of the $INTEGRAL$ bursts
GRB\,040223, GRB\,040422 and GRB\,040624 in the first hours after the triggers. A 
detection of a faint afterglow and of the host galaxy in the $K$ band was achieved 
for GRB\,040422, while only upper limits were obtained for GRB\,040223 and GRB\,040624,
although in the former case the X-ray afterglow was observed. A comparison 
with the magnitudes of a sample of afterglows clearly shows the faintness of these 
bursts, which are good examples of a population that an increasing usage of 
large diameter telescopes is beginning to unveil.
\end{abstract}

\maketitle


\section{INTRODUCTION}

The study of GRB afterglows is a promising tool for cosmology, since their 
absorption spectra convey information on the distance and the chemical 
composition of a new set of galaxies (e.~g. \cite{fiore}), with the possibility 
of exploration up to the reionization epoch \cite{lamb}. However, a debated 
fraction of GRBs -- from less than 10\% \cite{hete} to 
60\% \cite{lazzati} -- did not show any detectable afterglow in the optical 
band. Popular and non-mutually exclusive explanations are: these bursts have 
intrinsically faint afterglows in the optical band (e.~g. \cite{fynbo}; 
\cite{lazzati}); their decay is very fast \cite{berger}; the optical 
afterglow is extinguished by dust in the vicinity of the GRB or in the 
star-forming region in which the GRB occurs (e.~g. \cite{lamb2000}; 
\cite{reichart}); their redshift is above 6, so that the Lyman-$\alpha$ 
absorption by neutral hydrogen in the host galaxy and along the line of sight 
damps the optical radiation of the afterglow \cite{lamb2000}. To these 
physical explanations, one must add the possibility that the search techniques 
are neither accurate nor quick enough \cite{hete}. The possibility that some 
afterglows are intrinsically faint has of course a big impact on the
modelling of the GRBs themselves, as well as on their application in cosmology. On the 
other hand, if one or several of the other explanations are correct, a 
substantial reduction of the fraction of dark bursts can be achieved by quick 
observations in the infrared. In any case, a fast and multiwavelength follow-up 
campaign of observations is mandatory for the study of GRB afterglows. To this
aim, the ESA's International Gamma-Ray Astrophysics Laboratory \textit{INTEGRAL} 
\cite{winkler}, launched in October 2002, has a burst alert system called IBAS 
(\textit{INTEGRAL} Burst Alert System, \cite{Mere03}). 
IBAS carries out rapid localizations for GRBs incident on the IBIS detector 
with a precision of a few arcminutes \cite{mere:2004}. The public 
distribution of these coordinates enables multi-wavelength searches for 
afterglows at lower energies. \textit{INTEGRAL} data on the prompt emission in 
combination with the early multi-wavelength studies, such as those presented in 
this work for GRB\,040223, GRB\,040422 and GRB\,040624, can probe these high 
energy phenomena.

\section{The afterglow of GRB\,040422}

\begin{figure}
  \includegraphics[height=.3\textheight]{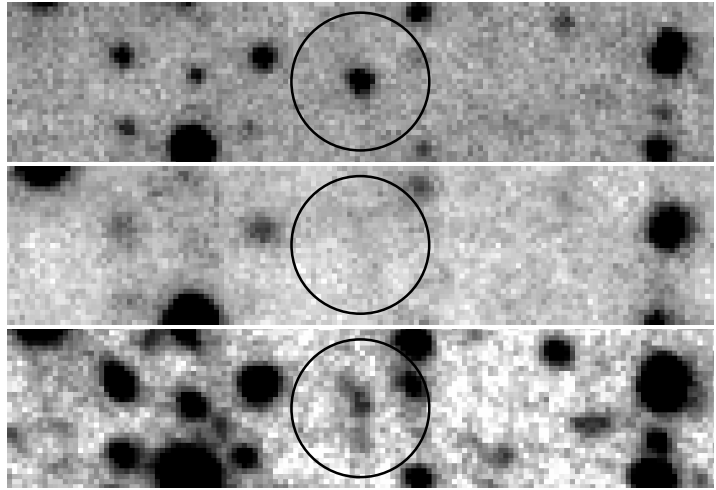}
  \caption{The region of the afterglow of GRB\,040422. From top to bottom: 2004
  Apr 22.37, 2004 May 05.33, 2004 Jun 26.11 UT.}
\label{fig:040422}
\end{figure}

GRB\,040422 was detected by the \textit{INTEGRAL} satellite at an angle of only 
3 degrees from the Galactic plane. We observed the 
afterglow of GRB\,040422 with the ISAAC and FORS\,2 
instruments at the VLT less than 2 hours after the burst. Such a prompt
reaction, together with careful inspection of the crowded field of this GRB, 
led to the discovery of its near-infrared afterglow, for which we obtained the 
astrometry and photometry. We measured for the afterglow a magnitude 
$K = 18.0 \pm 0.1$ (1.9 hours after the burst), a value which is below the limit 
of the 2MASS catalogue. 
No detection could be obtained in the $R$ and $I$ bands, 
partly due to the large extinction in the Milky Way. We imaged the position of 
the afterglow again two months later in the $K$ band, and detected a 
likely bright ($K \sim 20$) host galaxy (Fig.~\ref{fig:040422}, for more details
see~\cite{filliatre1}). We compare the magnitude of the afterglow with those 
of a compilation of promptly observed counterparts of previous GRBs, and show 
that the afterglow of GRB\,040422 lies at the very faint end of the 
distribution, after accounting for 
Milky Way extinction (Fig.~\ref{fig:pl}). This observation suggests that the 
proportion of dark GRBs can be lowered significantly by a more systematic use 
of 8-m class telescopes in the infrared in the very early hours after the burst. 

\section{The dark GRB\,040223 and GRB\,040624}

\begin{figure}
\includegraphics[height=.3\textheight]{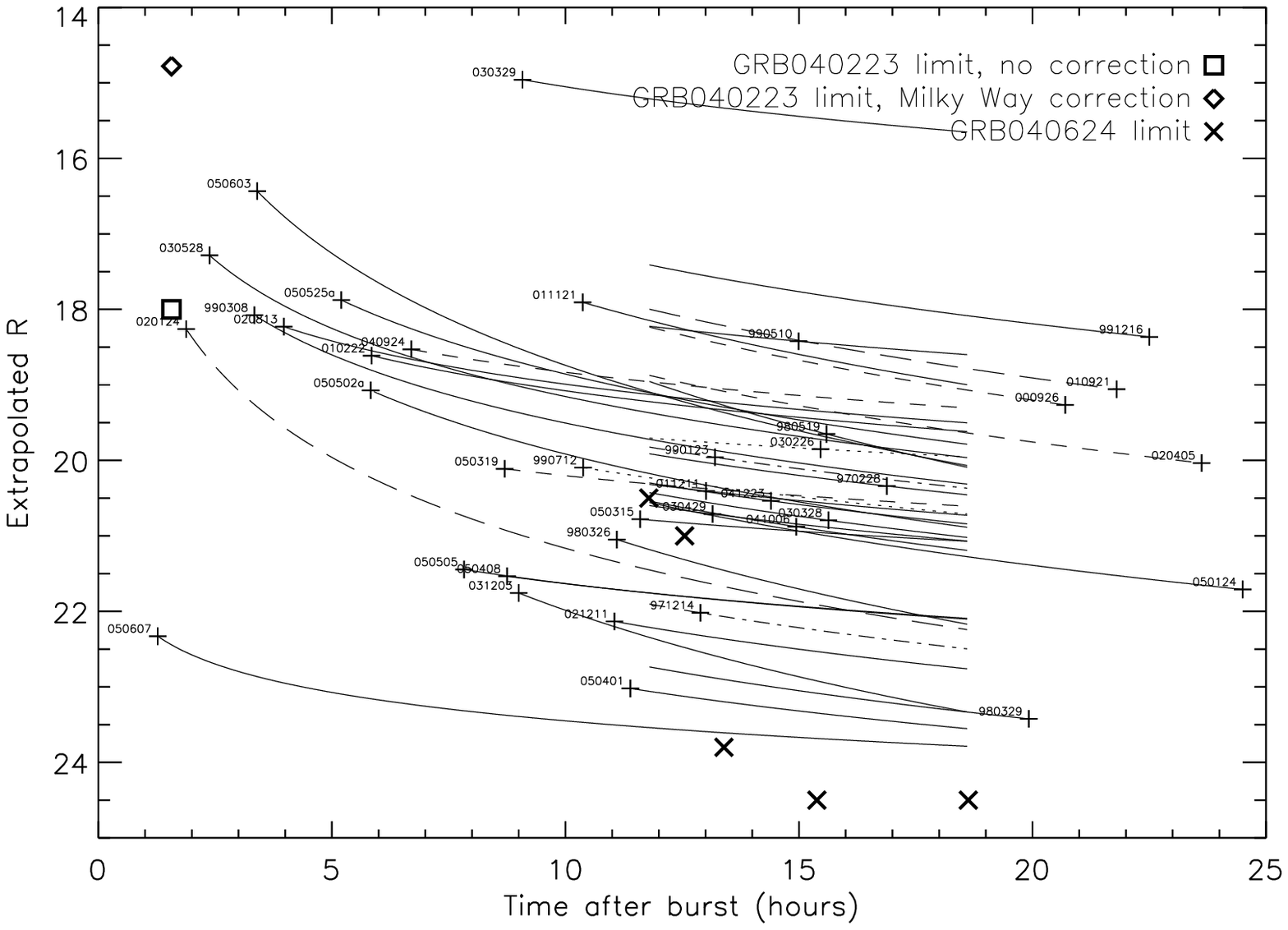}
\includegraphics[height=.3\textheight]{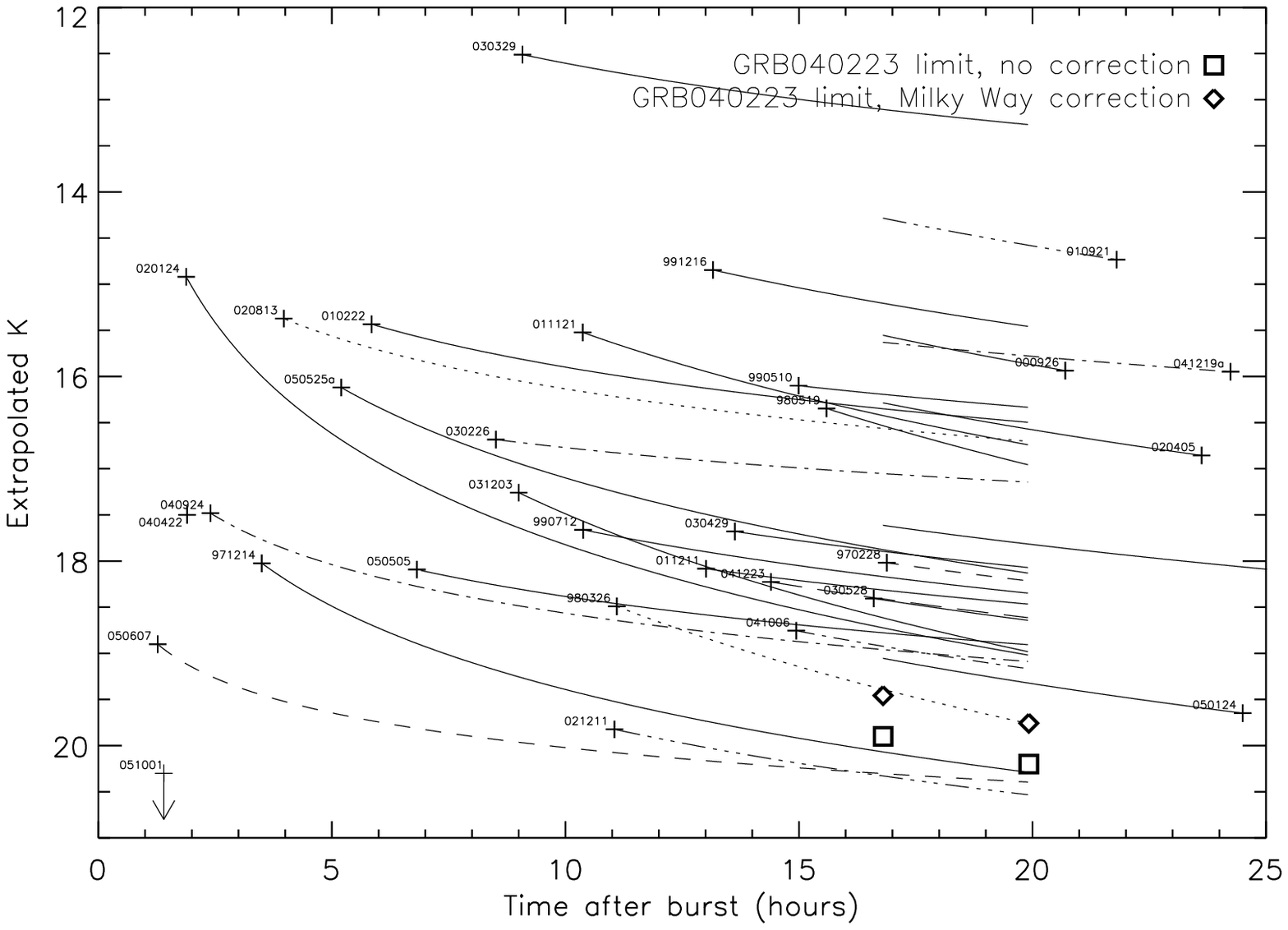}
\caption{Light curves of a set of 39 afterglow data, observations 
of GRB\,040422 and limits of GRB\,040223 and GRB\,040624.
For clarity, different line styles are used to indicate the temporal
power-law decay extrapolations, which only cover our
observation epochs.
($Left$) Magnitudes extrapolated to
the $R$ band (when necessary) after correction for Galactic
absorption. The cross is placed at the time of observation. The diamond and 
the square indicate the magnitude limits for {GRB\,040223} with and 
without correction for Galactic extinction, respectively, at
2-$\sigma$ (\cite{Gomb04}). The crosses indicate the magnitude
limits for {GRB\,040624}. The first two come from the GCNs
(\cite{Picc04,Goros04}). The last three points are the 3-$\sigma$ limits reported
in \cite{filliatre2}.
($Right$) Magnitudes extrapolated to
the $K$ band (when necessary) after correction for
Galactic absorption. The cross is placed at the time of
observation. The diamonds and the squares
indicate our 3-$\sigma$ magnitude limits with and without
correction from Galactic extinction, respectively. The $K$ magnitude for 
GRB\,040422 (\cite{filliatre1}) and the limit for GRB\,051001 (\cite{gcn4053}) 
are also reported.
}
\label{fig:pl}
\end{figure}

GRB\,040223 was detected by \textit{INTEGRAL} close to the Galactic plane while 
GRB\,040624 was at high Galactic latitude. The two GRBs have long durations, slow 
pulses and are weak. The $\gamma$-ray spectra of both bursts are
best fit with steep power-laws, implying they are X-ray rich. GRB\,040223 is 
among the weakest and longest of \textit{INTEGRAL} GRBs.
The X-ray afterglow of this burst was detected 10 hours after
the prompt event by \textit{XMM-Newton}. The measured spectral
properties are consistent with a  column density much higher
than that expected from the Galaxy, indicating strong
intrinsic absorption. We carried out near-infrared observations 17
hours after the burst with the ESO-NTT, which yielded upper
limits. Given the intrinsic absorption, we find that these limits
are compatible with a simple extrapolation of the X-ray afterglow
properties. For GRB\,040624, we carried out optical
observations 13 hours after the burst with FORS\,1 and 2 at the
VLT, and DOLoRes at the TNG, again obtaining upper limits.
As for GRB\,040422, we compare these limits with the magnitudes of a compilation
of promptly observed counterparts of previous GRBs and find again that
they lie at the very faint end of the distribution (Fig.~\ref{fig:pl}, for a more
detailed analysis see~\cite{filliatre2}). Together with GRB\,040422, 
these two bursts are good examples of a population of bursts with dark
or faint afterglows that are being unveiled through the
increasing usage of large diameter telescopes engaged in
comprehensive observational programmes.

%


\begin{thebibliography}{}

\bibitem{berger} 
Berger, E., Kulkarni, S. R., Bloom, J. S., et al. 2002, \emph{ApJ}, 581, 981

\bibitem{filliatre1} 
Filliatre, P., D'Avanzo, P., Covino, S., et al. 2005, \emph{A\&A}, 438, 793

\bibitem{filliatre2} 
Filliatre, P., Covino, S., D'Avanzo, P., et al. 2005, \emph{A\&A accepted, astro-ph/0511722}

\bibitem{fiore} 
Fiore, F., d'Elia, V., Lazzati, D., et al. 2005, \emph{ApJ}, 624, 853

\bibitem{fynbo} 
Fynbo, J. U., Jensen, B. L., Gorosabel, J., et al. 2001, \emph{A\&A}, 369, 373

\bibitem{Gomb04} 
Gomboc A., Marchant J.M., Smith R.J., Mottram C.J., Fraser S.N. 2004,
\emph{GCN\,2534}

\bibitem{Goros04} 
Gorosabel J., Casanova V., Verdes-Montenegro L., et al. 2004, \emph{GCN\,2615}

\bibitem{lamb2000} 
Lamb, D. Q. 2000, \emph{Phys. Rep.}, 505, 333

\bibitem{lamb} 
Lamb, D. Q., \& Reichart, D. E. 2000, \emph{ApJ}, 536, 1

\bibitem{hete} 
Lamb, D. Q., Ricker, J. R., Atteia, J.-L., et al. 2004, \emph{New A Rev.}, 48, 423

\bibitem{lazzati} 
Lazzati, D., Covino, S., \& Ghisellini, G. 2002, \emph{MNRAS}, 330, 583 

\bibitem{Mere03} 
Mereghetti, S., G\"{o}tz, D., Borkowski, J., Walter, R., \& Pedersen, H. 2003, 
\emph{A\&A}, 411, L291

\bibitem{mere:2004} 
Mereghetti, S., G\"{o}tz, D., Borkowski, J., et al. 2004a, in Proceedings of the
5th $INTEGRAL$ Workshop: The $INTEGRAL$ Universe (Munich), ESA 
Special Publication SP-552, ed. V. Sch\"onfelder, G. Lichti \& C. Winkler, 
\emph{astro-ph/0404019}

\bibitem{Picc04} 
Piccioni A., Bartolini C., Guarnieri A., et al. 2004, \emph{GCN\,2623}

\bibitem{reichart} 
Reichart, D. E. \& Price, P. A. 2002, \emph{ApJ}, 565, 174

\bibitem{gcn4053} 
Rol,  N., Levan, A., E., Tanvir, et al. 2005b, \emph{GCN\,4053}

\bibitem{winkler} 
Winkler, C., Courvoisier, T.~J., Di Cocco, G., et al. 2003, \emph{A\&A}, 411, L1

%
%
%
%

\end{thebibliography}
\end{document}